\documentclass[aps,reprint,superscriptaddress,twocolumn,nofootinbib]{revtex4}
\pdfoutput=1

\usepackage[intlimits]{amsmath}
\usepackage{amssymb}
\usepackage{mathrsfs}
\usepackage{slashed}
\usepackage{graphicx}
\usepackage{libertine}
\usepackage{microtype}
\usepackage{hyperref}

\usepackage[usenames,dvipsnames]{color}

\newcommand{\Dfb}{\mbox{$\raisebox{2mm}{${}^\leftrightarrow$}\hspace{-4mm} D_\mu$}}
\newcommand{\DfbI}{\mbox{$\raisebox{2mm}{${}^\leftrightarrow$}\hspace{-4mm} D^I_\mu$}}

\begin{document}

\title{Self-interacting neutrinos: solution to Hubble tension versus experimental constraints}

\author{Kun-Feng Lyu}
\email[]{klyuaa@connect.ust.hk}
\affiliation{Department of Physics, The Hong Kong University of Science and Technology, Clear Water Bay, Kowloon, Hong Kong S.A.R., P.R.C.}
\affiliation{Kavli Institute for Theoretical Physics, University of California, Santa Barbara, CA 93106, USA}

\author{Emmanuel Stamou}
\email[]{emmanuel.stamou@epfl.ch}
\affiliation{Theoretical Particle Physics Laboratory (LPTP), Institute of Physics, EPFL, Lausanne, Switzerland}

\author{Lian-Tao Wang}
\email[]{liantaow@uchicago.edu}
\affiliation{Enrico Fermi Institute, University of Chicago, Chicago, IL 60637, USA}
\affiliation{Kavli Institute for Cosmological Physics, University of Chicago, Chicago, IL 60637, USA}

\date{\today}

\begin{abstract}
Exotic self-interactions among the Standard-Model neutrinos have been proposed as
a potential reason behind the tension in the expansion rate, $H_0$, of the universe inferred
from different observations.
We constrain this proposal using electroweak precision observables,
rare meson decays, and neutrinoless double-$\beta$ decay.
In contrast to previous works, we emphasize the importance of carrying
out this study in a framework with full Standard-Model gauge invariance.
We implement this first by working with a relevant set of
Standard-Model-Effective-Field-Theory operators and subsequently
by considering a UV completion in the inverse See-Saw model.
We find that the scenario in which all flavors of neutrinos self-interact universally is
strongly constrained, disfavoring a potential solution to the $H_0$ problem
in this case.
The scenario with self-interactions only among
tau neutrinos is the least constrained and can potentially be
consistent with a solution to the $H_0$ problem.
\end{abstract}

\maketitle

\section{Introduction}

There is a tantalizing discrepancy between the value of the Hubble constant
($H_0$) extracted from local measurement versus the one extracted from the
Cosmic Microwave Background
data~\cite{Aghanim:2018eyx,Riess:2018kzi,Shanks:2018dsp,Riess:2019cxk,Wong:2019kwg}.

Towards this end, the authors of Ref.~\cite{Kreisch:2019yzn} suggested
to give neutrinos a new, extra strong self-coupling
in the form of the dimension-six operator
\begin{equation}
	\mathscr{L}_{\text{eff}} \sim G_{\nu} (\bar{\nu}_M\nu_M)(\bar{\nu}_M\nu_M)\,.
\label{eq:eff_coupling}
\end{equation}
We focus on the possibility that the Standard-Model (SM) neutrinos are Majorana, i.e., $\nu_M$ are four-component Majonara fermion fields.
The Dirac case is strongly disfavoured by
Big-Bang-Nucleosynthesis (BBN) constraints \cite{Blinov:2019gcj}.
This effective interaction can be induced by the presence
of a light scalar mediator --- a massive version of the so-called
Majoron\cite{Gelmini:1980re} --- with an effective coupling
\begin{equation}
	\mathscr{L}_{\text{eff Majoron}} \sim \lambda \phi \bar{\nu}_M \nu_M\,.
\label{eq:majoron}
\end{equation}
The effect of neutrino self-interactions in cosmological observables has been
investigated in
Refs.~\cite{Oldengott:2017fhy,Lancaster:2017ksf,Archidiacono:2013dua,Cyr-Racine:2013jua,Huang:2017egl}.
The interaction in Eq.~\eqref{eq:eff_coupling} can postpone
the time at which the neutrinos
begin to free stream and induce a phase shift towards high-$l$ scale at the CMB TT spectrum.
Together with one additional sterile neutrino, which brings
$N_{\text{eff}} \simeq 4$, this can reduce the tension in $H_0$.
The fit of the CMB data favours two values for $G_{\nu}$,
namely, SI$\nu$ (Strongly Interacting) and MI$\nu$  (Moderately Interacting)
\begin{equation}
G_{\nu} =
       \begin{cases}
	       (4.6(5)~\text{MeV})^{-2} &(\text{SI}\nu)\\
               (90^{+170}_{-60} ~\text{MeV})^{-2} &(\text{MI}\nu)
       \end{cases}\,.
       \label{eq:fitres}
\end{equation}

There have been many studies on the constraints of the neutrino--Majoron coupling.
Experimental results like Supernova~\cite{Farzan:2002wx,Kachelriess:2000qc},
neutrinoless double-$\beta$ decay~\cite{Gando:2012pj,Arnold:2018tmo},
Meson decays~\cite{Britton:1993cj,Lazzeroni:2012cx,NA62:2011aa,Lessa:2007up,Bakhti:2017jhm,Brdar:2020nbj} and $Z$-pole
observables~\cite{Abbiendi:2003dh,Brdar:2020nbj} all give relevant constraints,
see Ref.~\cite{Blinov:2019gcj} for a summary of various bounds in the strong
self-coupling scenario.
However, most of the studies focus on the effective neutrino--Majoron
coupling in Eq.~\eqref{eq:majoron}, which violates electroweak gauge invariance.
This is perfectly fine as long as one focuses on the degrees of freedom
well below the weak scale.
On the other hand, we have established a very accurate description of the physics
around the weak scale, known as the Standard Model (SM).
There are precision measurements that will set
relevant constraints on the scenario of self-interacting neutrinos.
Indeed, many of the studies did implemented such constraints, e.g.,
$Z$ decays.
It is now mandatory to go beyond the effective
interaction in Eq.~\eqref{eq:majoron} and consider weak-scale UV completions.
While our analysis here is motivated by the solution to the Hubble tension,
the results are general constraints on the
neutrino self-coupling, whether it would play a role in interpreting
the CMB data or not.

In this work, we take two consecutive steps in this direction.

Firstly, we will remain (mostly) agnostic about the specifics of new physics
and assume that, apart from the Majoron itself, it is somewhat
heavier than the weak scale.
Hence, we will parameterize their effect by dimension-five and six
effective operators in the Standard Model Effective Theory
(SMEFT).
One such dimension-six operator contains the Majoron and
induces neutrino self-interactions.
However, in typical models that modify the neutrino sector
and induce neutrinos masses the aforementioned operator
is accompanied by additional ones, which do not contain the Majoron,
and are typically generated in any UV completion.
We will use experimental data to constrain the
size of their Wilson coefficients.

Secondly, we will consider the possibility of UV completing this
effective theory
into renormalizable models by introducing new degrees of freedom.
Neutrinos are embedded in ${\rm SU}(2)$ doublets, therefore,
at the renormalizable level the neutrino--Majoron coupling can
only be induced via the (mass) eigenstate mixing after electroweak
symmetry breaking.
There are two paradigms: mixing with a neutrino sector or with a scalar sector.
The former is realized in the Type-I seesaw model
while the latter in Type-II.
In both cases, the mixing angle determines the strength of the
neutrino--Majoron coupling.
However, we will see that for Type-I, the mixing is
proportional to the neutrino mass and is thus too suppressed
to provide a sufficiently large mixing.
Similarly for Type-II, the current bound on the triplet Yukawa coupling and
the vev of the triplet scalar implies that it cannot provide a sufficiently
large mixing either\cite{Perez:2008ha,Cai:2017mow}.
However, we will show that there exist extended seesaw models in
which there is no direct connection between the mixing and the neutrino mass.
One of them is the so-called inverse seesaw model, which we will consider in detail.
We will match the model to the SMEFT operators, and use the constraints
derived for them to set limits on the model parameters.

We will find that within a SM gauge invariant framework,
the extent to which neutrino self-interactions may alleviate
the $H_0$ inconsistency depends on the flavour structure of
the self-couplings.
The case in which all flavours interact with the
same strength (universal) is too constrained from electron-sector
observables to provide a solution.
However, the case in which only tau-flavor neutrinos self-interact
may still provide a solution due to the weaker constraints from
particle-physics observations.

The rest of this paper is organized as follows:
In section~\ref{sec:framework}, we
describe the relevant SMEFT framework and match it to seesaw
models.
In section~\ref{sec:observables}, we present the
predictions for the observables entering the analysis.
In section~\ref{sec:fit} we combine the observables
and contrast them to the CMB fit and discuss the various
regions of the parameter space.
We conclude in section~\ref{sec:conclusions}.

\section{The framework\label{sec:framework}}
\subsection{Neutrino self-interactions within the extended SMEFT\label{sec:smeft}}

We begin with the assumption that, with the exception of the
Majoron $\phi$, new physics is heavier than the electroweak scale.
In this case, all beyond-the-SM effects can be parameterized by a set
of non-renormalizable operators.
In our case, we are interested in a small subset of operators that
induce neutrino self-interactions and those that typically accompany them
in UV-complete models.
More specifically, the following set suffices to capture the main
phenomenological aspects
\begin{multline}
	\label{eq:LEFT}
\mathscr{L}_{\text{EFT}} =   C^f_{\nu\nu} ~( Q^f_{\nu\nu} +\text{h.c.})
	                + C^f_\phi~\phi (Q^f_\phi+\text{h.c.})\\
			+ C^f_{ew}~ (Q_{HL}^{(1),f} - Q_{HL}^{(3),f})\,.
\end{multline}
$f$ denotes the neutrino flavor with $f=e,\mu,\tau$, and
\begin{equation}
	\label{eq:EFToperators}
	\begin{split}
	Q^f_{\nu\nu}   &= \overline{L^c}_f \tilde{H}^* \tilde{H}^\dagger L_f\\
	Q^f_\phi       &=\phi \overline{L^c}_f \tilde{H}^* \tilde{H}^\dagger L_f\\
	Q_{HL}^{(1),f} &= (H^\dagger i\,\Dfb H) (\overline{L}_f\gamma^\mu L_f)\\
	Q_{HL}^{(3),f} &= (H^\dagger i\,\DfbI H) (\overline{L}_f\sigma^I\gamma^\mu L_f)\,.
	\end{split}
\end{equation}
Our notation follows closely the ones of Ref.~\cite{Grzadkowski:2010es}.
We ignore flavor-changing operators and restrict the discussion
to flavor-diagonal operators.

The SM neutrinos live in the weak doublets $L_f$, thus the Higgs doublet $H$
must be included to form gauge singlets.
The dimension-five Weinberg operator, $Q^f_{\nu\nu}$, accounts
for neutrino masses.
The operator $Q^f_\phi $ is responsible for generating the self-interaction.
The operators $Q_{HL}^{(1),f} $ and $Q_{HL}^{(3),f}$ must also be included,
because they are typically also generated at the tree-level
in models that induce $Q^f_{\nu\nu}$.
In particular, the operators $Q_{HL}^{(1),f}$ and $Q_{HL}^{(3),f}$
are typically generated  with a Wilson coefficient
of same magnitude but opposite sign, i.e.,
$C_{HL}^{(1),f} = C_{ew}^f$ and
$C_{HL}^{(3),f} = -C_{ew}^f$
(we have already implemented this in Eq.~\eqref{eq:LEFT}).
The reason behind this tree-level relation is that typical models that
generate the $Q^f_{\nu\nu}$ operator by integrating out some heavy degrees
of freedom, also necessarily induce the derivative operator
$(\overline{L}_f \tilde{H}) i\slashed{\partial} (\tilde{H}^\dagger L_f)$.
This derivative operator is redundant in the Warsaw basis, where it is
removed in favor of the combination
$Q_{HL}^{(1),f} - Q_{HL}^{(3),f} = 4 (\overline{L}_f \tilde{H}) i\slashed{\partial} (\tilde{H}^\dagger L_f)$.
The presence of these operators lead to important phenomenological consequences,
which cannot be captured when one simply works with the effective coupling
in Eq.~\eqref{eq:majoron}.

To work with dimensionless couplings for the dimension-six Wilson coefficients
we introduce the notation $\bar{C}_X= C_X v^2$, with $v \simeq 246$ GeV the
electroweak vev.

At scales below the electroweak scale the $Q^f_{\nu\nu}$ operators
induce a Majorana mass term for the neutrinos,
and the $Q^f_{\phi}$ couplings of the neutrinos to $\phi$.
The resulting Lagrangian reads
\begin{equation}
	\mathscr{L}_{\nu} = \frac{1}{2} \bar\nu_{M,f} (i\slashed{\partial} - m_{\nu_f})\nu_{M,f} + \frac{1}{2} \lambda_f \phi \bar\nu_{M,f}\nu_{M,f}
	\label{eq:LbelowEW}
\end{equation}
with $\nu_{M,f} = \nu_{L,f} + (\nu_{L,f})^c$ the four-component Majorana fermion
and where
\begin{align}
	&m_{\nu_f} = -\bar C^f_{\nu\nu}\,,& &\lambda_f = \bar C^f_{\phi}\,.&
\end{align}
We note that both the mass and the interaction in Eq.~\ref{eq:LbelowEW}
are flavor diagonal.
We emphasize that this is an assumption, and more general flavor structures are
certainly possible.
However, the aim of this work is to extract main lessons rather than carry
out an exhaustive study.
Moreover, as we will see in section~\ref{sec:Gnu}, the effect of
neutrino self-interactions on the CMB has only been studied
under a quite (over)simplified case.
Hence, we will also make simplifying assumptions for the
flavor structure in our study.

\subsection{Seesaw Models}
\subsubsection{Type-I Seesaw Model}
To illustrate how the EFT operators presented in section~\ref{sec:smeft}
are induced in concrete UV models we start with the
simplest Type-I seesaw model.
The SM Lagrangian is augmented with an extra heavy right-handed neutrino
\begin{multline}
	\mathscr{L} \supset  \overline{N_R} i \slashed{\partial} N_R
	-\dfrac{M_R}{2}(\overline{N_R^c}N_R + \overline{N_R}N_R^c)\\+ (-y_R \bar{L} \tilde{H} N_R  + \dfrac{1}{2}\lambda \phi \overline{N_R^c} N_R + \text{h.c.})\,.
\end{multline}
with $N_R$ a four-component chiral field.
After electroweak symmetry breaking, the mixed Dirac mass is generated $m_D = y_R v/\sqrt{2}$.
The neutrino mass matrix then reads
\begin{equation}
M = \begin{pmatrix}
0 & m_D\\
m_D & M_R
\end{pmatrix}\,.
\end{equation}
After diagonalization,
the masses of the light mass eigenstates in the limit $m_D \ll M_R$ are
\begin{equation}
	m_\nu = \frac{m_D^2}{M_R}\,,
\end{equation}
and the mixing angle between light and heavy eigenstates is
\begin{equation}
	\sin \theta \sim \dfrac{m_D}{M_R}\,.
\end{equation}
Hence, the coupling between the Majoron and light eigenstates reads
\begin{equation}
	g_{\phi \nu \nu} = \lambda\left(\dfrac{m_D}{M_R}\right)^2 = \lambda\dfrac{m_\nu}{M_R}\,.
\end{equation}
We see that in this model the coupling to the Majoron is
suppressed by the neutrino mass and thus cannot produce
strong self-interactions for perturbative values of $\lambda$.

To match to the effective Lagrangian in Eq.~\eqref{eq:LEFT}, we
integrate out $N_R$ at the tree-level and find the Wilson coefficients
\begin{align}
	&C^f_{\nu\nu}    = - \frac{y_R^2}{2} \frac{1}{M_R},&
	&C^f_\phi        = \frac{\lambda}{2}\frac{y_R^2}{M_R^2},&
	&C^f_{ew}        = \frac{1}{4}\frac{y_R^2}{M_R^2}.&
	\label{eq:type-i-coeff}
\end{align}
Again, we see that $C^f_{\nu\nu}$, which generates the neutrino mass,
is correlated to $C^f_\phi$.
Hence, the neutrino--Majoron interaction, proportional to $C^f_\phi$,
is suppressed by the neutrino mass.

\subsubsection{Inverse Seesaw Model\label{sec:modelinvseesaw}}
In order to break the correlation between $C^f_{\nu\nu}$ and $C^f_\phi$
we consider an inverse seesaw model
\cite{Dias:2012xp,Law:2013gma,Mohapatra:1986bd,Ma:1987zm,Ma:2009gu,Bazzocchi:2010dt}
augmented with an additional real scalar, $\phi$, that couples
to one species of the heavy neutrinos:
\begin{align}
	\mathscr{L}_{\text{inv-seesaw}} &=
	i \overline{\mathcal{F}} \slashed{\partial} \mathcal{F}
	- M \overline{\mathcal{F}} \mathcal{F} \nonumber\\
&-\left(
       \frac{\delta_R}{2} \overline{\mathcal{F}_R^c} F_R
      + \frac{\delta_L}{2} \overline{\mathcal{F}_L^c} F_L
      + y_{R,f} \overline{L}_f \tilde{H} \mathcal{F}_R
      + \text{h.c.}\right)\nonumber\\
&+\frac{\lambda}{2}\phi\left(\overline{\mathcal{F}_L^c} F_L
      + \text{h.c.}\right)\,,
      \label{eq:invseesaw}
\end{align}
with $\mathcal{F} = \mathcal{F}_L+\mathcal{F}_R$.
The fermion fields $\mathcal{F}_L$ and $\mathcal{F}_R$
above are four-component chiral fields, i.e., only two components are non-zero.
By choosing to couple the Majoron only to $\mathcal{F}_L$ and not to
$\mathcal{F}_R$ we break the correlation between neutrino mass and
Majoron coupling. The subscript, $f$,  stands for the flavor. For simplicity
we consider the heavy-neutrino setting for each flavor separately
and do not consider their mixing.

We match to the effective Lagrangian in Eq.~\eqref{eq:LEFT}
by integrating out the heavy fields $\mathcal{F}_R$ and $\mathcal{F}_L$
at the tree-level.
For the case $\delta_L,\,\delta_R \ll M$ and up to dimension-six the Wilson
coefficients we obtain are
\begin{align}
	&C^f_{\nu\nu}    = - \frac{y_{R,f}^2}{2} \frac{\delta_L}{M^2}\,,&
	&C^f_\phi        =   \frac{\lambda}{2}\frac{y_{R,f}^2}{M^2}\,,&
	&C^f_{ew}        =   \frac{1}{4}\frac{y_{R,f}^2}{M^2}\,.&
	\label{}
\end{align}
Contrary to the Type-I model, we see that the neutrino mass and the
neutrino--Majoron coupling are controlled by independent parameters,
$\delta_L$ and $\lambda$, respectively.

It is thus possible to induce a sizable neutrino--Majoron coupling without it
being suppressed by the neutrino mass.
At the same time, we see that $C^f_\phi$ and $C^f_{ew}$ are correlated to some extent,
which has important phenomenological consequences.

To include constraints from electroweak-precision observables,
we also compute the Wilson coefficient of the operator that
contributes to the $T$-parameter at tree level.
In the Warsaw basis this operator is $Q_{HD} \equiv |H^\dagger D^\mu H|^2$.
The one-loop matching at a scale $\mu\simeq M$ gives
\begin{equation}
	L \supset C_{HD} Q_{HD}\quad\text{with}\quad C_{HD}(M) = - \frac{1}{16\pi^2} \frac{y_{R,f}^4}{2 M^2}\,,
	\label{<+label+>}
\end{equation}
where we only kept terms of ${\cal O}(y_{R,f}^4)$.
We include the leading terms of ${\cal O}(y_{R,f}^2 g_1^2, y_{R,f}^2 y_e^2)$
by solving the renormalization group (RG) within SMEFT
(see section~\ref{sec:Tparam}).

\section{Observables\label{sec:observables}}
In this section we discuss the most relevant observables
in our analysis and their predictions within the SMEFT framework.
We choose as the numerical input for the electroweak parameters $G_F$, $\alpha$, and $m_Z$.
As extensively discussed in the literature, e.g., Ref.~\cite{Brivio:2017btx} and references
within, the presence
of dimension-six operators affects the determination of the electroweak-parameter input.
In the case at hand, only the operators in Eq.~\eqref{eq:EFToperators} are induced at the
tree-level and in fact out of them only the operators $Q_{HL}^{(3),e}$ and $Q_{HL}^{(3),\mu}$
affect the extraction of $G_F$.
The remaining electroweak input remains unchanged.
The $G_F$ shift affects all electroweak observables.
To take it into account, one substitutes, e.g., Ref.~\cite{Brivio:2017btx},
\begin{multline}
G_F \longrightarrow G_F \left(1- \bar{C}_{HL}^{(3),e} - \bar{C}_{HL}^{(3),\mu}\right)=\\
	 = G_F \left(1+ \bar{C}_{ew}^{e} + \bar{C}_{ew}^{\mu}\right)\,,
	\label{<+label+>}
\end{multline}
where $G_F$ is still the experimental input value.

\subsection{$Z$ decays}
After electroweak-symmetry breaking the operator combination
$Q^{(1),f}_{HL}- Q^{(3),f}_{HL}$ does not (directly) affect the charged lepton sector, but
it does induce an anomalous $Z$ coupling to the neutrino species $f$, i.e.,
\begin{equation}
	\mathscr{L}_{\text{anom-}Z} = -\frac{e}{2s_wc_w} 2 \bar C^f_{ew} \bar\nu^f_L\slashed{Z}\nu^f_L\,.
\end{equation}
Together with the shift in $G_F$ this modifies the partial width to the neutrinos
\begin{align}
	\Gamma(Z\to\bar\nu_f\nu_f) = \frac{G_F m_Z^3}{12\sqrt{2}\pi} (1 + \bar C^e_{ew} + \bar C^\mu_{ew} - 4 \bar C^f_{ew})\,.
	\label{<+label+>}
\end{align}

We also include the three-body partial width $Z\to \bar\nu_f\nu_f\phi$, which is, however,
formally higher order in the EFT, i.e., it is proportional to $(\bar C^f_\phi)^2$.
For the region of interest $m_\phi\ll m_Z$, we find for a neutrino species
coupled to $\phi$ via the operator $Q^f_\phi$ the width
\begin{equation}
	\Gamma(Z\to\bar\nu_f\nu_f\phi) = \frac{G_F m_Z^3}{12\sqrt{2}\pi}
	\frac{(\bar C^f_\phi)^2}{192\pi^2}
		\left(
			12\log\biggl(\dfrac{m_Z^2}{m_\phi^2}\biggr)
			-23
		\right)\,.
\end{equation}
Notice that this rate diverges for $m_\phi\to 0$. For small $m_\phi$ it is thus
necessary to resum the logarithms. However, for the masses that we are
considering this is not necessary.
Also due to the double EFT suppression this rate is numerically small.

Similarly we evaluate the effect of the shift in $G_F$ in the partial
width to charged leptons and to hadrons.
In the SM the partial width to a fermion $f$ with charge $Q^f$ is
\begin{equation}
	\Gamma(Z\to\bar f f)^{\text{SM}} = n_c^f \frac{\alpha m_Z}{24 s_w^2 c_w^2}\left( 1- 4s_w^2 |Q^f| + 8 s_w^4 |Q^f|^2 \right)\,.
	\label{<+label+>}
\end{equation}
After shifting $G_F$ we find that
\begin{multline}
	\Gamma(Z\to\ell^+\ell^-)   = \Gamma(Z\to\ell^+\ell^-)^{\text{SM}}\times\\\times
	\left( 1 +
	(\bar C_{ew}^e+\bar C_{ew}^\mu) \frac{1-2 s_w^2-4 s_w^4}{(1-2s_w^2)(1-4 s_w^2+8 s_w^4)}\right)\,,
\end{multline}
\begin{multline}
	\Gamma(Z\to\text{hadrons}) = \Gamma(Z\to\text{hadrons})^{\text{SM}}\times\\\times
	\left( 1 +
	(\bar C_{ew}^e+\bar C_{ew}^\mu) \frac{45-90 s_w^2-4 s_w^4}{(1-2s_w^2)(45-84 s_w^2 + 88 s_w^4)}\right)\,.
\end{multline}

\subsection{$T$-parameter\label{sec:Tparam}}
Heavy sterile neutrinos can affect electroweak-precision observables, i.e., the $T$-parameter.
Within the SMEFT framework the new-physics contributions to the $T$-parameter
are controlled by the Wilson coefficient of the $Q_{HD}$ operator
evaluated at the electroweak scale, $\mu_{\text{ew}}\sim m_Z$,
via $\alpha  T = \hat{T} = -\frac{v^2}{2} C_{HD}(\mu_{\text{ew}})$
\cite{Barbieri:2004qk}.
In our setup we integrate out the heavy degrees of freedom at a scale
$M\gg \mu_{\text{ew}}$ and obtain $C_{HD}(\mu_{\text{ew}})$ via the RG
evolution to $\mu_{\text{ew}}$ (see Refs.~\cite{Alonso:2013hga,Jenkins:2013wua}
for the corresponding anomalous dimensions).
Operators that have been induced at the tree-level at $M$ can mix into $Q_{HD}$.
In our case we find that at leading-log accuracy
\begin{align}
C_{HD}(m_Z) &= C_{HD}(M)\\
            &-\frac{e^2}{16\pi^2} \left(\frac{8}{3c_w^2} + \frac{4}{s_w^2c_w^2}\frac{m_f^2}{m_Z^2}\right) C^{(1),f}_{HL}\log\frac{m_Z}{M}\,.\nonumber
\end{align}
The singlet operators $Q_{HL}^{(1),f}$ mix into $Q_{HD}$,
introducing the dependence on $C^{(1),f}_{HL} = C_{ew}^f$.

\subsection{Leptonic meson decays}
Non-standard neutrino interactions affect the decays of pseudoscalar mesons.
The most stringent constraints originate from the semileptonic decays
of charged pseudoscalars.
The modification with respect to the SM originates both from the shift in $G_F$ and the
anomalous coupling $W\ell_f\nu_f$ proportional to $\frac{e}{\sqrt{2}s_w}\bar C^{(3),f}_{HL}$.
The two-body partial width of a pseudoscalar, $P$, to a neutrino and a charged lepton
then reads
\begin{multline}
\Gamma(P \rightarrow \ell_f\nu_f) = \dfrac{G_F^2}{8\pi} f_{P}^2 m_{\ell_f}^2 m_P
\Big( 1-\dfrac{m_{\ell_f}^2}{m_P^2}\Big) V_{\text{CKM}}\times\\\times
\left( 1 + 2 \big( \bar C^{e}_{ew} +\bar C^{\mu}_{ew}-\bar C^{f}_{ew}\big) \right)\,,
	\label{<+label+>}
\end{multline}
with $f_P$ the decay constant of the meson and $V_{\text{CKM}}=|V_{u_id_j}|^2$ the corresponding
CKM elements. As in the SM the two-body widths are helicity suppressed and
thus proportional to the charged lepton mass.

The helicity suppression is lifted in the three-body decay $P\rightarrow \ell_f\nu_f\phi$.
Expanding in the mass of the charged lepton we find
\begin{multline}
\Gamma(P \rightarrow \ell_f \nu_f \phi) = \dfrac{f_P^2 G_F^2 V_{\text{CKM}} m_P^3}{768 \pi^3}
({\bar C}_{\phi}^f)^2\times\\\times
\left(1+9 x_\phi-9 x_\phi^2-x_\phi^3 +6 (x_\phi+1) x_\phi \log x_\phi\right)\,,
\end{multline}
where $x_{\phi}=m_\phi^2/m_P^2$.

\subsection{Neutrino self-scattering ($G_{\nu}$)\label{sec:Gnu}}

The presence of new, neutrino self-interactions can modify the neutrino
standard free-streaming behavior during the radiation-dominated era.
$2\to 2$ scattering among neutrinos modifies the momentum dependence
of the neutrino distribution functions and can thus affect
cosmological observables such as the CMB.
The cosmological fit of Ref.~\cite{Kreisch:2019yzn} is performed for the
case in which the $\phi$ mass is much larger than the typical energy scale of
the scattering event.
In this case we can to an excellent approximation integrate out $\phi$ and
describe the neutrino self-interactions via four-fermion contact interactions.

Starting from the (per assumption) {\itshape flavor-diagonal} Lagrangian
for the four-component Majorana fermions $\nu_{M,i}$
in Eq.~\eqref{eq:LbelowEW} we use
the EOM of the real scalar ($(\Box+m_\phi^2)\phi = \frac{1}{2}\sum_i \bar C_\phi^i \bar\nu_{M,i}\nu_{M,i}$)
to obtain the effective Lagrangian
\begin{equation}
	\begin{split}
	\mathscr{L}_{\nu,\text{eff}}
	&= \frac{1}{8m_{\phi}^2}\sum_{i,j} \bar C_\phi^i \bar C_\phi^j (\bar\nu_{M,i}\nu_{M,i})(\bar\nu_{M,j}\nu_{M,j})\\
	&= \frac{1}{8}\sum_{i=1,2,3}                        C_{\nu\nu}^{i}  (\bar\nu_{M,i}\nu_{M,i})(\bar\nu_{M,i}\nu_{M,i})\\
	&+
	   \frac{1}{4}\sum_{\substack{i,j=1,2,3\\i < j}} C_{\nu\nu}^{ij} (\bar\nu_{M,i}\nu_{M,i})(\bar\nu_{M,j}\nu_{M,j})
   \end{split}
	\label{eq:Lnueff}
\end{equation}
with
\begin{align}
	&C_{\nu\nu}^{i}  = \frac{(\bar C_\phi^i)^2}{m_\phi^2}\,,&
	&C_{\nu\nu}^{ij} = \frac{\bar C_\phi^i \bar C_\phi^j}{m_\phi^2}\quad\text{with}~i<j\,.&
	\label{<+label+>}
\end{align}
Here, the indices $i,j=1,2,3$ indicate the flavors $e,\mu,\tau$, respectively.
Note that when $\phi$ couples flavour-diagonally to more that one flavour
a mixed four-fermion operator is necessarily generated.

In order to make contact with the results of the CMB fit of Ref.~\cite{Kreisch:2019yzn}
we present here the corresponding collision terms for neutrino scattering.
In the general, flavour-diagonal case there are three independent processes:
the self-scattering of one species ($\nu_i + \nu_i \to \nu_i + \nu_i$),
$s$-channel annihilation ($\nu_i + \nu_i \to \nu_j + \nu_j$ with $i\neq j$),
and $t$-channel scattering ($\nu_i + \nu_j \to \nu_i + \nu_j$ with $i\neq j$).
Their respective squared matrix-elements summed over initial- and final-state spins are:
\begin{align}
|{\cal M}^i_{s,t,u}|^{2} &\equiv  \sum_{\text{spins}}|{\cal M}_{\nu_i\nu_i\to \nu_i\nu_i}|^2\nonumber\\
&= 2 (C_{\nu\nu}^i)^2\left( s^2+t^2+u^2 \right)\nonumber\\
&= 2 \frac{(\bar C_{\phi}^i)^4}{m_\phi^4} \left( s^2+t^2+u^2 \right) \label{eq:M2iiii}\,,\\
|{\cal M}^{ij}_s|^{2} &\equiv \sum_{\text{spins}}|{\cal M}_{\nu_i\nu_i\to \nu_j\nu_j}|^2 = 4 (C_{\nu\nu}^{ij})       s^2\nonumber\\
&= 4 \frac{(\bar C_{\phi}^i)^2(\bar C_{\phi}^j)^2}{m_\phi^4}) s^2 \qquad\text{with}\quad i<j\,,\\
|{\cal M}^{ij}_t|^{2} &\equiv \sum_{\text{spins}}|{\cal M}_{\nu_i\nu_j\to \nu_i\nu_j}|^2
= 4 (C_{\nu\nu}^{ij})^2 t^2\nonumber\\
&= 4 \frac{(\bar C_{\phi}^i)^2(\bar C_{\phi}^j)^2}{m_\phi^4}) t^2 \qquad\text{with}\quad i<j\,,
\end{align}
with $s,t,u$ the usual Mandelstam variables.
No symmetry factors for identical particles have been included above.

What enters the evolution of the neutrino distributions
are the collision integrals for each process.
Adapting the generic expression from Ref.~\cite{Kolb:1990vq}
we find that for a specific neutrino species $i$ and $j\neq i$ the
collision integrals for the three processes above are:
\begin{multline}
	\mathscr{C}_{\nu_i(p_1)\nu_i(p_2)\leftrightarrow \nu_i(p_3)\nu_i(p_4)}
	= \frac{1}{2g}\int d\Pi_2 d\Pi_3 d\Pi_4~(2\pi)^4\times\\
	\times F[\nu_i(p_1),\nu_i(p_2); \nu_i(p_3), \nu_i(p_4)]\times\\
	\times \delta(p_1+p_2-p_3-p_4)~|{\cal M}^i_{s,t,u}|^2\,,\label{eq:col1}
\end{multline}
\begin{multline}
	\mathscr{C}_{\nu_i(p_1)\nu_i(p_2)\leftrightarrow \nu_j(p_3)\nu_j(p_4)}
	= \frac{1}{2g}\int d\Pi_2 d\Pi_3 d\Pi_4~(2\pi)^4\times\\
	\times F[\nu_i(p_1),\nu_i(p_2); \nu_j(p_3), \nu_j(p_4)]\times\\
	\times\delta(p_1+p_2-p_3-p_4)~\frac{1}{2}|{\cal M}^{ij}_{s}|^2\,,\label{eq:col2}
\end{multline}
\begin{multline}
	\mathscr{C}_{\nu_i(p_1)\nu_j(p_2)\leftrightarrow \nu_i(p_3)\nu_j(p_4)}
	= \frac{1}{2g}\int d\Pi_2 d\Pi_3 d\Pi_4~(2\pi)^4\times\\
	\times F[\nu_i(p_1),\nu_j(p_2); \nu_i(p_3), \nu_j(p_4)]\times\\
	\times \delta(p_1+p_2-p_3-p_4)|{\cal M}^{ij}_{t}|^2\,,\label{eq:col3}
\end{multline}
where $	d\Pi_i = \frac{d^3p_i}{(2\pi)^32E_i}$ and $F[\dots]$
defined as in Ref.~\cite{Kreisch:2019yzn}.
The factor $1/g$, with $g=2$ the spin degrees of freedom, has been omitted in
Ref.~\cite{Kreisch:2019yzn}.
No additional symmetry factors for identical particles in initial and final state
need to be included in Eq.~\eqref{eq:col1} (see Ref.~\cite{Gondolo:1990dk}).
The factor $1/2$ in Eq.~\eqref{eq:col2} is due to the identical particles $j\neq i$ in the
final or initial state.

We see that the collision integrals in Eqs.~\eqref{eq:col2} and \eqref{eq:col3}
couple the evolution of the distribution function of the three species.
This cross-talk has been been neglected in Ref.~\cite{Kreisch:2019yzn}.
Instead each neutrino flavour was assumed to self-interact independently with
the same strength and the fit to the CMB provided the best-fit value for the
parameter $G_{\nu}$ defined via the collision integral
for each species \cite{Kreisch:2019yzn}
\begin{multline}
	\mathscr{C}^{\text{Ref.~\cite{Kreisch:2019yzn}}}_{\nu_i(p_1)\nu_i(p_2)\leftrightarrow \nu_i(p_3)\nu_i(p_4)}
	= \frac{1}{2}\int d\Pi_2 d\Pi_3 d\Pi_4~(2\pi)^4\times\\
	\times F[\nu_i(p_1),\nu_i(p_2); \nu_i(p_3), \nu_i(p_4)]\\
	\times \delta(p_1+p_2-p_3-p_4)~2~G_{\nu}^2(s^2+t^2+u^2)\,,\label{eq:col4}
\end{multline}
We emphasize again that this is an over-simplifying assumption that
does not follow from flavor universality.
Nevertheless, we will use it since it provides a direct comparison
between the explicit fit performed in Ref.~\cite{Kreisch:2019yzn} and
the constraints obtained in this paper.
Under this simplifying assumption, i.e., neglecting cross-talk,
we find by comparing Eqs.~\eqref{eq:col1} and \eqref{eq:col4}
that for the {\itshape ``universal case''} ($C_{\nu\nu}^{1} = C_{\nu\nu}^{2} = C_{\nu\nu}^{3} =
C_{\nu\nu}^{12} = C_{\nu\nu}^{13}= C_{\nu\nu}^{23}\equiv C_{\nu\nu}$)
\begin{equation}
	C_{\nu\nu} \longleftrightarrow \sqrt{2} G_{\nu} \,.
	\label{eq:relation}
\end{equation}
Additionally to the ``universal'' case, we also consider
{\itshape ``flavor specific''} cases
(one $C^i_{\nu\nu}\neq 0$ and all other couplings zero)	in which
the self-interactions take place only among a single species
instead of among all three.
These cases are governed by the evolution of the thermal bath of
one neutrino with the collision integral in Eq.~\eqref{eq:col1}.
The fit of Ref.~\cite{Kreisch:2019yzn} does not cover these cases,
a dedicated re-analysis is required, which is beyond the scope of the present
work.
Roughly, the total strength of self-interactions are weaker
if a single species self-interacts than in the ``universal case''.
To at least partially take this into account we interpret the fit
of Ref.~\cite{Kreisch:2019yzn} for the {\itshape ``flavor specific''} case via the rescaling
\begin{equation}
	C^i_{\nu\nu} \longleftrightarrow \sqrt{6} G_{\nu} \,.
	\label{eq:relationspecific}
\end{equation}
The results of a future CMB fit for these cases could then be obtained by a simple rescaling of
Eq.~\eqref{eq:relationspecific}.
We caution that while we expect this scaling to partially take into account
the difference between the effective coupling strength,
the factor of $\sqrt{6}$ is just an educated guess.
More complete numerical study is needed to obtain the precise factor.

For the cases we consider, the two best-fit regions from Ref.~\cite{Kreisch:2019yzn}
in Eq.~\eqref{eq:fitres} then translate into:
\begin{multline}
	v^2 |C_\phi| = \frac{m_\phi}{\text{MeV}} ~
	\times
       \begin{Bmatrix}
	       (4.6(5))^{-1} &(\text{SI}\nu)\\
	       (90^{+170}_{-60})^{-1} &(\text{MI}\nu)
       \end{Bmatrix}
       \times\\\times
       \begin{Bmatrix}
	       2^{\frac{1}{4}} & (\text{``flavor-specific'' case})\\
	       6^{\frac{1}{4}} & (\text{``universal'' case})\\
       \end{Bmatrix}\,.
       \label{eq:Gnunum}
\end{multline}

\section{Numerical Analysis\label{sec:fit}}

As we have discussed above, we focus on two distinct limits
in both of which the self-interactions via $\phi$ are assumed to be aligned to the neutrino
mass-eigenstates.
\begin{description}
	\item[``Flavor-specific'' cases:] Self-interactions are present only for one, the $f$-th species
		of neutrinos with $f=e, \mu,\tau$. In this case
		$\bar{C}_{\phi}^f\,,\bar{C}_{ew}^f\neq 0$ while
		$\bar{C}_{\phi}^{f'} = \bar{C}_{ew}^{f'} = 0$ for $f' \neq f$.
	\item[``Universal'' case:] All three neutrinos species interact with equal strength such that
			$\bar{C}_\phi \equiv \bar{C}_{\phi}^e =\bar{C}_{\phi}^\mu = \bar{C}_{\phi}^\tau$ and
			$\bar{C}_{ew} \equiv \bar{C}_{ew}^e = \bar{C}_{ew}^\mu = \bar{C}_{ew}^\tau$.
\end{description}

\subsection{Experimental input / Constraints}
To illustrate the relative importance of various particle-physics and cosmological observables
in the ``flavor-specific'' and ``universal'' cases
we perform $\chi^2$ fits combining information from multiple observables.
Below we summarize the experimental input relevant for the fits.
Any additional, unspecified numerical
input is taken from Ref.~\cite{Tanabashi:2018oca}.
\begin{enumerate}
\item {\it Z decays:}
	We implement the constraints from the partial width
	measurements of the $Z$ boson by
	centering the corresponding $\chi^2$'s around the SM predictions and using the
	experimental uncertainties \cite{Tanabashi:2018oca}
\begin{align*}
	\Delta \Gamma_{\ell^+ \ell^-} &= 0.086 \text{ MeV}\,,\\
	\Delta \Gamma_{\text{had}} &= 2.0 \text{ MeV}\,,\\
	\Delta \Gamma_{\text{inv}} &= 1.5 \text{ MeV}\,.
\end{align*}
\item{\it $T$-parameter:}
When discussing the inverse seesaw model we also include the constraint from the $T$-parameter
as it can be affected by the presence of heavy neutrinos.
We use the current best fit value of $T= 0.06 \pm 0.06$ \cite{Tanabashi:2018oca}.
\item{\it Meson decays:}
	Analogously to $Z$ decays also for meson decays we assume that the experimental
	measurements of their branching ratios and their lifetimes are centered around
	their SM predictions and add the corresponding experimental
	uncertainties in their $\chi^2$.
	We neglect subleading theory
	uncertainties associated to
	form-factors.
	We consider constraints from branchings fractions of two-body leptonic decays of
	$\pi^+$, $K^+$, $D_s^+$, as well as their lifetimes \cite{Tanabashi:2018oca}:
	\begin{align*}
		\Delta \text{BR}(\pi^+\to e^+\nu, \mu^+\nu)
		&= 4\!\times\! 10^{-7}\,,4\!\times\! 10^{-7}\\
		\Delta \text{BR}(K^+\to e^+\nu, \mu^+\nu)
		&= 7\!\times\! 10^{-8}\,,1.1\!\times\! 10^{-3}\,,\\
		\Delta \text{BR}(D_s^+\to \mu^+\nu,\tau^+\nu)
		&= 2.3\!\times\! 10^{-4}\,,2.3\!\times\! 10^{-3}\,,\\
		\Delta \tau_{\pi^+}
		&= 5\!\times\! 10^{-12}\,\text{sec}\,,\\
		\Delta \tau_{K^+}
		&= 2\!\times\! 10^{-11}\,\text{sec}\,,\\
		\Delta \tau_{D_s^+}
		&= 4\!\times\! 10^{-15}\,\text{sec}\,.
	\end{align*}
	Note that often measurement of ratios of branching fractions
	are more constraining than those from the branching ratios above.
	However, using such ratios can leave certain directions unconstrained
	when more than one neutrino species self-interact, i.e., in the ``universal'' case.
	The combination of constraints are, however, similar when folded with the
	lifetimes measurements and $Z$ decays.
	Therefore, to enable a better comparison between different cases
	we do not include ratios of branching ratios in the fits.
\item{\it Neutrinoless double $\beta$-decay:}
	As discussed in Ref.~\cite{Blum:2018ljv},
	current neutrinoless double $\beta$-decay experiments like NEMO-3~\cite{Arnold:2013dha} and
	KamLAND-Zen~\cite{Gando:2012pj} can stringently constrain light $e$-flavor Majorons.
	This will be illustrated by mapping the results of Figure 4 of Ref.~\cite{Blum:2018ljv}
	into our corresponding exclusion plots.
\item{\it BBN:}
  Strong constraints are imposed on light species remaining in thermal equilibrium
  with neutrinos as extra relativistic degrees of freedom during the BBN period,
  as they affect the effective number of neutrinos, $\Delta N_{\text{eff}}$.
  Here, we follow the analysis of Ref.~\cite{Blinov:2019gcj} and consider the mass
  of the real scalar to be above $1$\,MeV.
\end{enumerate}

\subsection{SMEFT fit\label{sec:SMEFTfit}}

\begin{figure*}[t]
	\centering
	\includegraphics[width=0.9\textwidth]{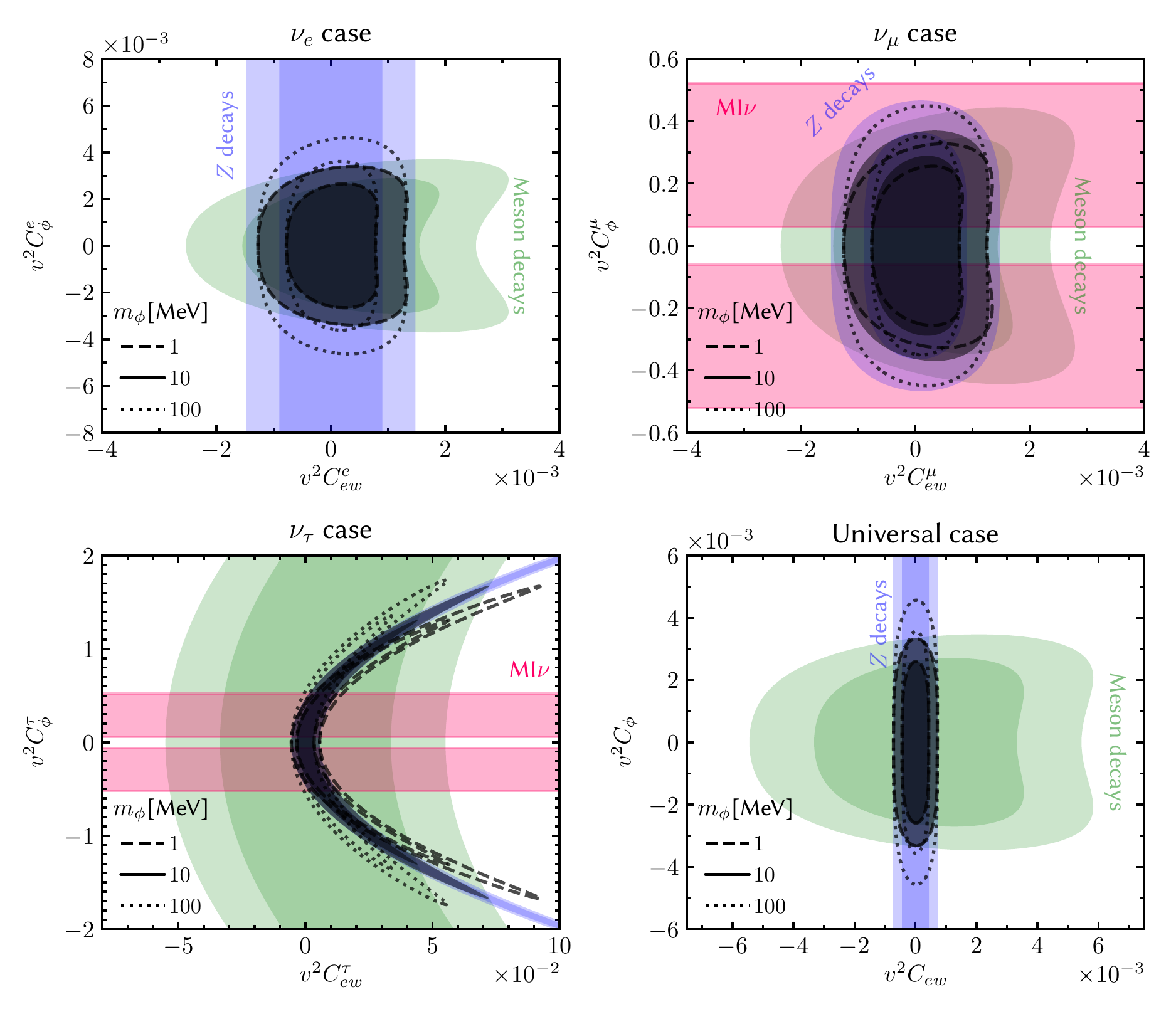}
	\caption{Preferred $68.27\%$ and $95.45\%$ CL regions for the Wilson coefficients $\bar{C}^{(i)}_{ew}$ and $\bar{C}_{\phi}^{(i)}$.
	Each panel corresponds to one of the four cases ($\nu_e$, $\nu_\mu$, $\nu_\tau$, and universal).
	In purple and green the constraints from $Z$ and leptonic meson decays, respectively.
	In black (gray) the combined allowed $68.27\%$ ($95.45\%$) CL region.
	The red regions in the second and third plot
	correspond to the $1\sigma$ preferred region
	for MI$\nu$ in Ref.~\cite{Kreisch:2019yzn}, cf., Eq.~\eqref{eq:Gnunum}.
	The best-fit regions for the SI$\nu$ case and the
	MI$\nu$ case not appearing in the first and last plot lie outside the ranges.
	All coloured regions correspond to $m_\phi=10$\,MeV.
	For the combined constraints we show the allowed region for  $m_\phi=1$\,MeV and $m_\phi=100$\,MeV
	in dashed and dotted lines, respectively.
	\label{fig:SMEFTfit}}
\end{figure*}

We first investigate the constraints on the SMEFT Wilson coefficients
for the four different cases ($e$, $\mu$, $\tau$, universal)
without specifying a UV model.
In each case there are three independent parameters
$\bar{C}_{ew}^{(i)}$, $\bar{C}_\phi^{(i)}$, and $m_\phi$.

In figure~\ref{fig:SMEFTfit}, we consider the four cases and
show the allowed $68.27\%$ and $95.45\%$ CL regions
for the two Wilson coefficients.
The purple and green regions are the allowed regions from $Z$ and mesons
decays, respectively, for the case $m_\phi=10$~MeV.
The black and gray regions are the combined allowed regions.
The dashed lines enclose the allowed region for $m_\phi=1$~MeV
and the dotted ones the region for $m_\phi=100$~MeV.
We also show the best-fit regions for the
strength of neutrino self-interactions from
Ref.~\cite{Kreisch:2019yzn}, cf., Eq.~\eqref{eq:Gnunum} when they lie
within the plot ranges.

Inspecting figure~\ref{fig:SMEFTfit} we observe that:
\begin{itemize}
\item The constraints from $Z$ decays (blue) and meson decays (green)
	are often complementary, e.g., in the $\nu_e$ case.
\item The main difference between the three ``flavor-specific'' cases
	are the constraints from meson decays. They are strongest for the
	$\nu_e$ case (top-left plot) and rather weak for the $\nu_\tau$ case
	(bottom--left plot).
	The reason is the different
	helicity suppression of the two-body meson decays,
	phase-space, and the fact that the $\nu_\tau$ case
	is only constrained by $D_s^+\to\tau^+\nu$.
	In contrast, the $\nu_e$ and $\nu_\mu$ cases receive strong
	constraints from $\pi^+$ and $K^+$ decays to $e^+ \nu_e$ and $\mu^+ \nu_\mu$.
\item The ``universal'' case (bottom-right plot) is to a large extent controlled by the its
	$\nu_e$ component and is thus similarly stringently constrained as
	the $\nu_e$ case.
\item The particle physics constraints on the
	$\nu_e$ and ``universal'' cases cannot be accommodated in
	neither the SI$\nu$ nor the MI$\nu$ best-fit regions of Ref.~\cite{Kreisch:2019yzn}
	for $m_\phi>1$\,MeV.
\item The MI$\nu$ best-fit regions (red) are compatible with particle-physics
	constraints for the $\nu_\mu$ and $\nu_\tau$ cases.
	Note, however, that the corresponding values for $\bar C_\phi$
	are ${\cal O}(1)$ thus close to the validity region of the EFT.
\end{itemize}

\subsection{Inverse seesaw model\label{sec:invseesawfit}}

\begin{figure*}[]
	\centering
	\includegraphics[width=0.85\textwidth]{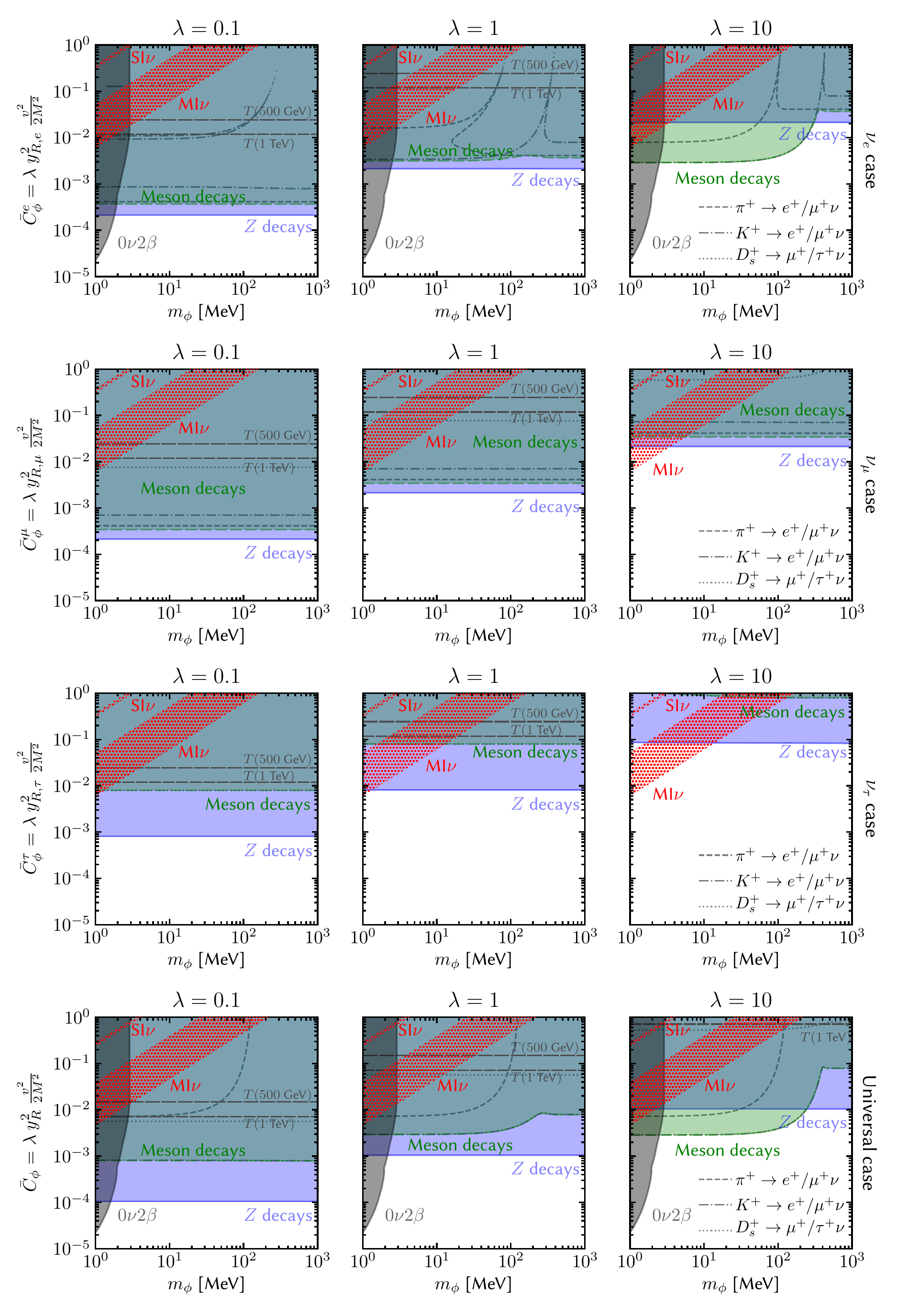}
	\caption{
		Exclusion regions for various cases of the inverse seesaw model
		in the $\bar C_{\phi}^{(f)}-m_\phi$ plane.
First-, second-, and third-row plots correspond to the flavor-specific
$e$-, $\mu$-, and $\tau$-case, respectively.
Fourth-row plots correspond to the ``universal'' case.
Each column shows the case of different values of $\lambda=0.1,1,10$.
The colored regions are excluded at $90\%$ CL: in purple the combined constraints
from $Z$ decays, in green the combined constraints from meson decays, and in grey
the constraints from neutrinoless double-$\beta$ decay \cite{Blum:2018ljv}.
Dashed lines indicated in the legend show the constraints from each meson sector
separately.
The red-dotted regions are the preferred $1\sigma$ regions of the CMB fit.
The horizontal, dashed lines show the constraint from the $T$-parameter when
the heavy-neutrino scale is $M=500$\,GeV and $M=1$\,TeV.}
	\label{fig:inverseseesawfit}
\end{figure*}

In the previous section we considered the particle-physics
constraints in conjunction with the preferred region from the CMB fit within
the mostly model-independent framework of SMEFT.
In concrete models, the SMEFT Wilson coefficients can be correlated, reducing
the number of free parameters and leading to correlated signals.
To illustrate this, we now study the phenomenology of the
inverse-seesaw model from section~\ref{sec:modelinvseesaw}.
Similarly to before we consider separately the three ``flavor-specific''
cases and the ``universal'' one.
In each case, we vary the $\phi$ mass and the effective Majoron coupling to
neutrinos, $\bar C_\phi^{(f)}$, while keeping the UV coupling $\lambda$ fixed.
As representative values for $\lambda$ we take $\lambda=0.1,\,1,\,10$.
We consider the case  $m_\phi > 1$\,MeV.
Smaller values of $m_\phi$ are constrained by BBN \cite{Ng:2014pca,Blinov:2019gcj}.

In figure~\ref{fig:inverseseesawfit}, we show the resulting constraints in the
$\bar C_{\phi}^{(f)}-m_\phi$ plane.
First-, second-, and third-row plots correspond to the flavor-specific
$e$-, $\mu$-, and $\tau$-case, respectively.
Plots of the fourth row correspond to the ``universal'' case.
Plots of each column present the case of different values of $\lambda$.
The colored regions are excluded at $90\%$ CL: in purple the combined constraints
from $Z$ decays, in green the combined constraints from meson decays, and in grey
the constraints from neutrinoless double-$\beta$ decay \cite{Blum:2018ljv}.
Dashed lines indicated in the legend show the constraints from each meson sector
separately, i.e., from $\pi^+$, $K^+$, and $D_s^+$ decays.
The red-dotted regions are the preferred $1\sigma$ regions of the CMB fit.
The horizontal, dashed lines show the constraint from the $T$-parameter when
the heavy-neutrino scale is $M=500$\,GeV and $M=1$\,TeV.

By inspecting figure~\ref{fig:inverseseesawfit} we recover some of the conclusions
from the SMEFT analysis of the previous section.
\begin{itemize}
\item The best-fit regions of the CMB fit cannot be accommodated
	in the ``flavor-specific'' $\nu_e$ and ``universal'' cases.
\item While the SI$\nu$ scenario is strongly disfavoured, the
	particle-physics constraints are compatible
	with the MI$\nu$ scenario in the ``flavor-specific'' $\nu_\mu$
	and $\nu_\tau$ cases, but only for masses $m_\phi\lesssim 10$\,MeV
	and large values of $\lambda$, i.e., $\lambda \gtrsim 1$, close
	to its perturbativity limit.
	This in turn implies that this scenario must have a cut-off
	close to the mass scale of exotic fermions.
\item The non-trivial structure of the $\pi^+$ (dashed-dotted lines) and $K^+$ (dashed lines)
	constraints in the $\nu_e$ and ``universal'' case is due to the interplay between
	the two-body decays, which suppresses the branching ratio BR$(M\to\ell\nu (\phi))$,
		and the three-body decay, which enhances it.
\item The scenario is being further tested at colliders by searches for the
      heavy neutrinos.
      The analyses, for example
      Refs.~\cite{Aad:2015xaa,Sirunyan:2018mtv}, typically search for the heavy-neutrino decays
      to $W$s and either electrons or muons, thus placing limits on the mass
      of the heavy neutrino for the flavor specific $e$ and $\mu$ cases, and
      not the $\tau$ case. In both $e$ and $\mu$ case, the present limits are
      rather weak, i.e., $M\gtrsim 100$\,GeV
      \cite{Aad:2015xaa,Sirunyan:2018mtv} for a mixing of the order
      $10^{-2}-10^{-3}$ between light and heavy neutrinos.
\end{itemize}
Qualitatively the results of this section are similar to
\cite{Ng:2014pca,Blinov:2019gcj}, but there are important differences.
In particular, the constraints from $Z$ decays, which are dictated by gauge invariance,
provide powerful constraints.
They restrict the allowed parameter-space of the $\nu_\tau$ ``flavor-specific'' case
more than meson decays.
The allowed region corresponds to large couplings, close to their perturbativity
bound.

\section{Conclusions\label{sec:conclusions}}

Motivated by the approach of using neutrino self-interactions to address
the tension in the $H_0$ measurement, we investigated the experimental
constraints on this scenario. In contrast to previous studies on this
setup, we began with an effective-field-theory framework that respects the full
Standard Model gauge symmetry. This is important as many of the
constraints are from experiments performed around the electroweak scale,
where the effect of electroweak symmetry is essential.
In addition to the SMEFT framework, we have also considered a UV
completion within an inverse-seesaw type model.
We performed an careful derivation of the constraints from $Z$ decay,
$T$-parameter, and meson decays. We also took into account the
limits from  the search of neutrinoless double-$\beta$ decay and BBN.
The constraints depends on the flavor structure of the couplings.
To illustrate this, we considered two scenarios. In one of them, the
self-interaction act in a ``flavor universal'' way to all flavors of
neutrinos. In the other one, there is only interaction between
one specific flavor species.

We showed that, in the ``flavor universal'' case, the neutrino
self-interaction as a solution to the $H_0$ problem is strongly disfavored.
Only the ``flavor-specific'' $\nu_\mu$ and $\nu_\tau$ cases in the
MI$\nu$ scenario may be provide a solution.
However, the scalar mass must be low  and the scalar--neutrino couplings large, close
to their perturbativity limits.
The SI$\nu$ scenario is strongly disfavoured.

Future experimental searches are promising in further testing these
scenarios. The experimental measurements  considered in this
paper will be improved significantly at on-going and future facilities.
The scenarios under consideration also point to new particles, for
example the new heavy neutrinos, not far away from the weak scale.
They can be searched for directly in the upcoming LHC runs and at
potential higher-energy colliders.

\section{Acknowledgements}
We would like to thank Sam McDermott and Massimiliano Lattanzi for useful discussions.
LTW is supported by the DOE grant DE-SC0013642.
KFL is supported in part by the Heising-Simons Foundation, the Simons Foundation,
and National Science Foundation Grant No. NSF PHY-1748958 and acknowledges the
hospitality of the Enrico Fermi Institute, where this work was initiated.
ES is supported by the Fermi
Fellowship at the Enrico Fermi Institute and by the U.S.\ Department
of Energy, Office of Science, Office of Theoretical Research in High
Energy Physics under Award No.\ DE-SC0009924 and
by the Swiss National Science Foundation under contract 200021--178999.

\bibliography{references}
\end{document}